\documentclass[aps,prl,preprint,groupedaddress,floatfix]{revtex4}

\usepackage{graphicx,latexsym}

\begin{document}


\title{On the Coalescence of Nanoscale Metal Clusters}


\author{S. Hendy}
\affiliation{Applied Mathematics, Industrial Research Ltd, Lower Hutt, New Zealand}
\altaffiliation{Also at MacDiarmid Institute for Advanced Materials
and Nanotechnology, School of Chemical and Physical Sciences, Victoria University of Wellington, New Zealand}
\author{S. A. Brown and M. Hyslop}
\affiliation{Nanostructure Engineering Science and Technology (NEST) Group and
MacDiarmid Institute for Advanced Materials and Nanotechnology, Department of Physics and Astronomy,
University of Canterbury, New Zealand}



\date{\today}

\begin{abstract}
We study the coalescence of nanoscale metal clusters in an inert-gas atmosphere using constant-energy molecular 
dynamics. The coalescence proceeds via atomic diffusion with the release of surface energy raising the temperature.
If the temperature exceeds the melting point of the coalesced cluster, a molten droplet forms. If the temperature
falls between the melting point of the larger cluster and those of the smaller clusters, a metastable molten droplet
forms and freezes.

\end{abstract}


\maketitle


The ability to predict and control the size and structure of nanoparticles is of great importance for applications,
as the arrangement of atoms is a key determinant of the optical, electronic and thermal properties of such particles.
However, the structure of an atomic cluster often differs from that of the corresponding bulk material \cite{ino66}.
As the number of surface atoms is comparable to the number of interior atoms in a cluster, the surface
energy plays an important role in determining the overall structure. The delicate balance between surface and internal
energies often produces a complex dependence of equilibrium structure upon cluster size \cite{Raoult89,Uzi97}.

In an inert-gas aggregation (IGA) source, where metal clusters are produced by evaporation and then
condensation from a vapor in an inert gas atmosphere, experiments have shown that the final cluster structure depends
on the growth rate \cite{Reinhard97}. Molecular dynamics simulations of silver clusters support this
finding, with studies of both the growth of clusters from small seeds \cite{Baletto00} and the freezing of clusters
from molten droplets \cite{Baletto02} finding a dependence of structure on kinetic factors. These examples illustrate
that it is important to understand cluster kinetics if one wishes to predict cluster structure.

The growth of clusters by coalescence is a kinetic process which will also effect structure. Indeed, the
aggregation of atomic clusters has generated much interest due to the variety of different large-scale structures that
can form, and the relevance of this to the functionality of devices built from clusters. For example, the deposition
of clusters onto surfaces can result in highly ramified fractal structures \cite{Brechignac01}, highly symmetric
droplets \cite{Goldby96}, as well as many intermediate structures \cite{Kaiser02}.  A recent proposal
\cite{Schmelzer02} uses cluster deposition to form wire-like cluster chains resembling nanowires with conduction of
electrical current between neighboring particles. Understanding cluster coalescence is therefore crucial for reliable
design and fabrication of such devices.

Further, the distribution of cluster sizes produced in IGA sources suggests that coalescence is an important growth
mechanism \cite{Granqvist76a,Granqvist76b}. Studies by electron diffraction of lead clusters grown in an IGA source
have found structures that could not be fitted by standard model structures but which might be accounted for by
coalesced structures \cite{Hyslop01}. To predict the structures of metal clusters produced in IGA sources we need to
study the coalescence of free metal clusters.

The collision and coalescence of small clusters, containing tens of atoms, has been studied extensively using molecular
dynamics. For example, an early study showed that the decrease in surface area during fusion can lead to an
increase in kinetic energy \cite{Schmidt91}. Simulations of the coalescence of larger clusters have been
predominantly conducted at constant temperature \cite{Zhu96,Lewis97}, which probably best represents the
coalescence of supported clusters with the substrate acting as a thermostat \cite{Lewis97}. However, as we will
argue below, the coalescence of nanometer-sized free clusters in an IGA source is best described by constant energy
molecular dynamics. We are aware of only one such study, which considered the fusion of both pairs of liquid and
pairs of "glassy" solid silicon clusters \cite{Zachariah99}, and demonstrated that the kinetic energy
generated during coalescence can considerably shorten coalescence times by enhancing atomic diffusion. Recent kinetic
Monte Carlo simulations have shown, however, that the coalescence of facetted clusters can be considerably
slower than surface-roughened particles \cite{Jensen00}.


Our purpose is to study the coalescence of facetted solid lead clusters using molecular dynamics under conditions
similar to those found in an IGA source. We examine the coalescence of lead clusters using the empirical glue
potential developed by Lim, Ong and Ercolessi \cite{LOE92}. Recently it was shown that many-body effects in the glue
potential produce unusual surface-reconstructed icosahedra in simulations of freezing \cite{Hendy01}. Further, these
surface-reconstructed icosahedra are thought to be globally optimum structures for the glue-potential for lead over a
wide range of sizes \cite{Hendy01,Doye03,Hendy02}. It is of interest then to study the coalescence of such icosahedra
under conditions that might occur in an IGA source.

An estimate of the mean time, $\tau_g$, for a cluster of radius $R$ between collisions with an inert-gas
atmosphere of atomic mass $m_g$, pressure $P$ and temperature $T$ in the source can be made as follows \cite{Baletto02}:
\begin{equation}
\tau_g \sim {1 \over P R^2} \sqrt{{m_g k T \over 8 \pi}} \nonumber.
\end{equation}
For a 2 nm radius lead cluster in a helium atmosphere at 5 mbar and 500 K, the time between collisions is approximately
1 ns. Thus over periods of a nanosecond or so, a constant-energy simulation of cluster coalescence will be
a reasonable approximation to conditions in an IGA source. We will use constant-energy molecular dynamics here.

Two 565-atom surface-reconstructed icosahedral clusters are initially prepared by construction as detailed in Hendy and
Doye \cite{Hendy02}. Each of these clusters consist of a 147 Mackay icosahedral core with two reconstructed outer shells
which differ from the usual Mackay icosahedral surface termination.
Each cluster is then equilibrated in isolation during a constant-temperature simulation for $10^5$ time-steps at some
common initial temperature $T_i$. The clusters are then placed in a common simulation cell so that the edges are aligned
and the corresponding edge atoms are 0.5 nm apart(approximately next-nearest neighbor distance but within the glue
potential cut-off). In this work, we will not examine the effects of relative orientation, initial structure or impact
parameter of the clusters in detail. Further simulations reveal that changing relative orientations does not alter the
conclusions reached here, although we have not investigated the case of non-zero impact parameter.

\begin{figure}
\resizebox{\columnwidth}{!}{\includegraphics{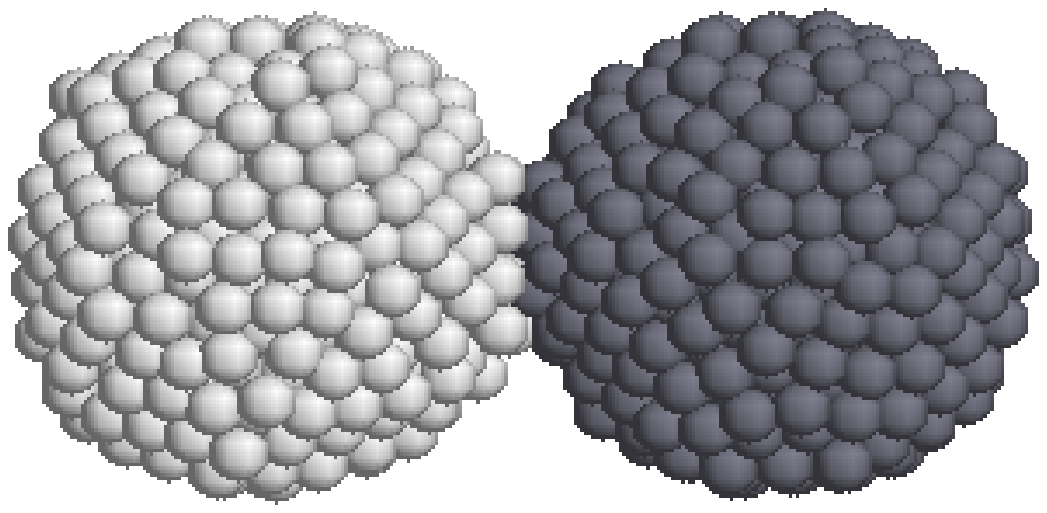} \includegraphics{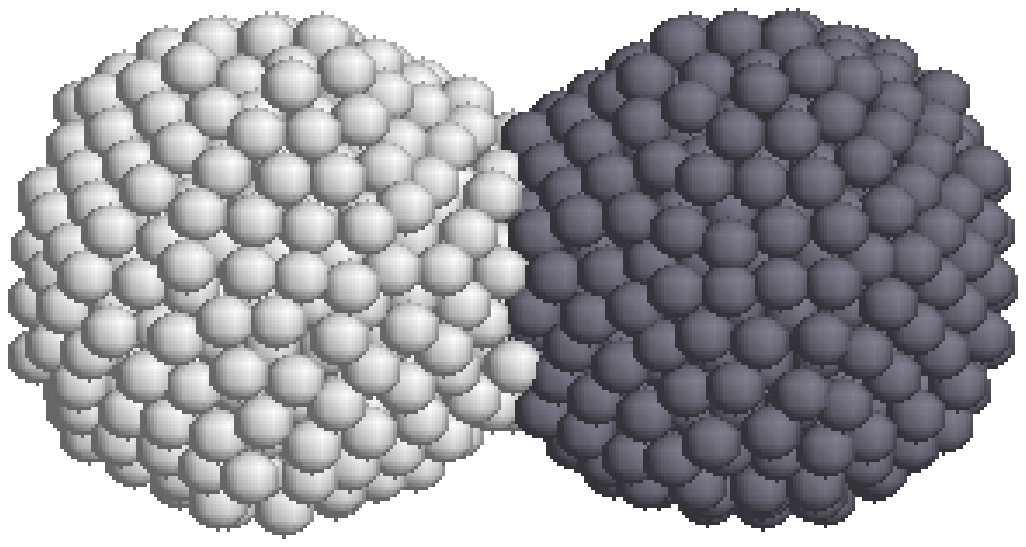} \includegraphics{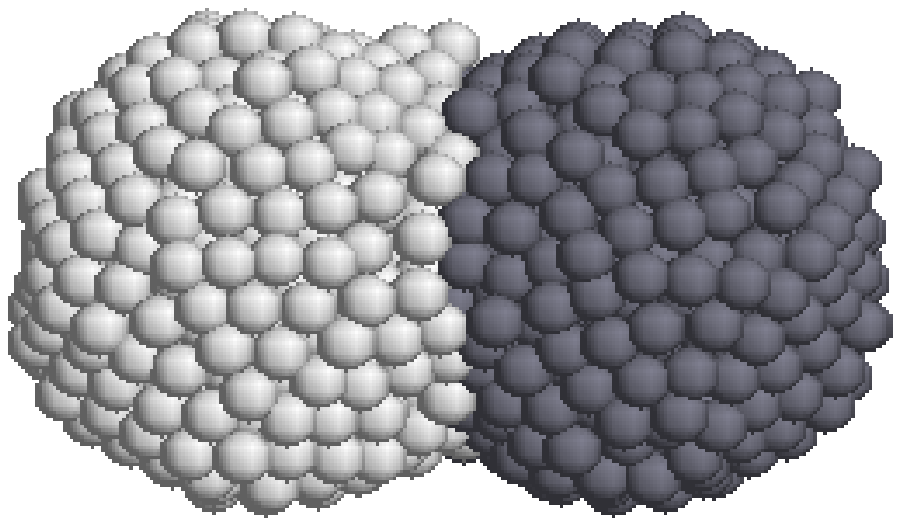}}
\resizebox{\columnwidth}{!}{\includegraphics{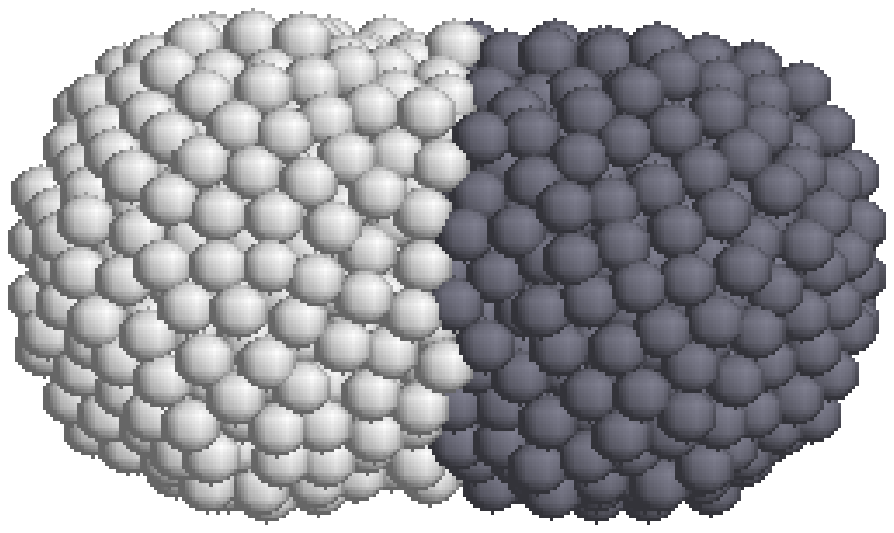} \includegraphics{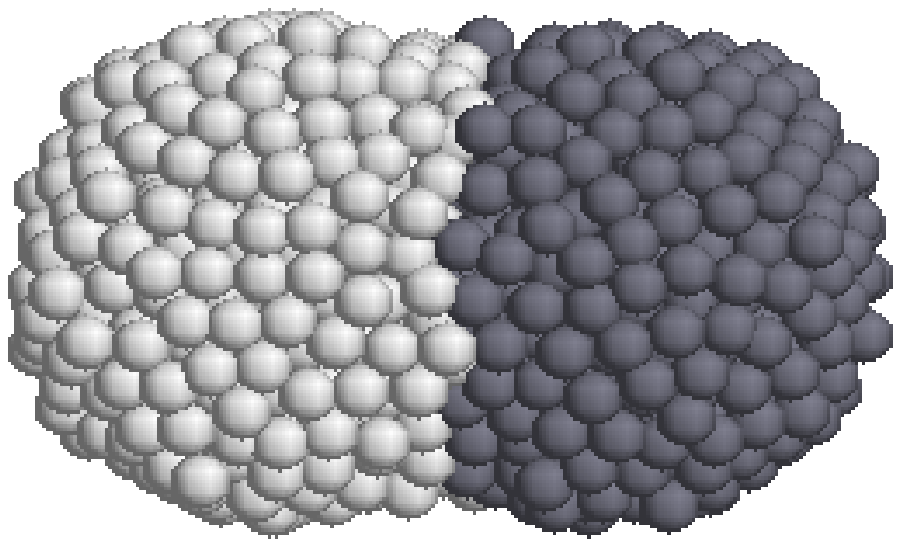} \includegraphics{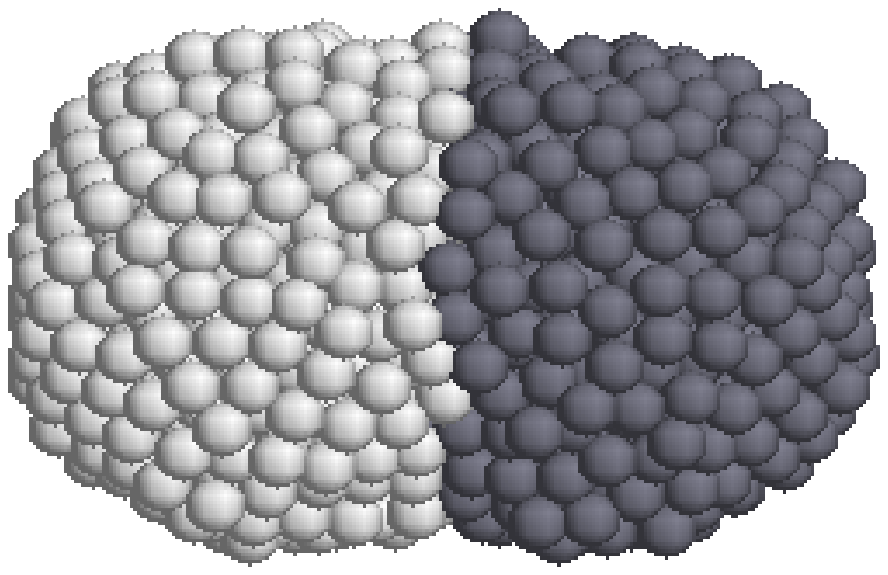}}
\caption{\label{frames} Coalescence of two 565-atom icosahedra initially at 300 K. The sequence of snapshots at $3.75$ ps
intervals shows the early growth of the neck after the initial contact.}
\end{figure}

\begin{figure}
\resizebox{\columnwidth}{!}{\includegraphics{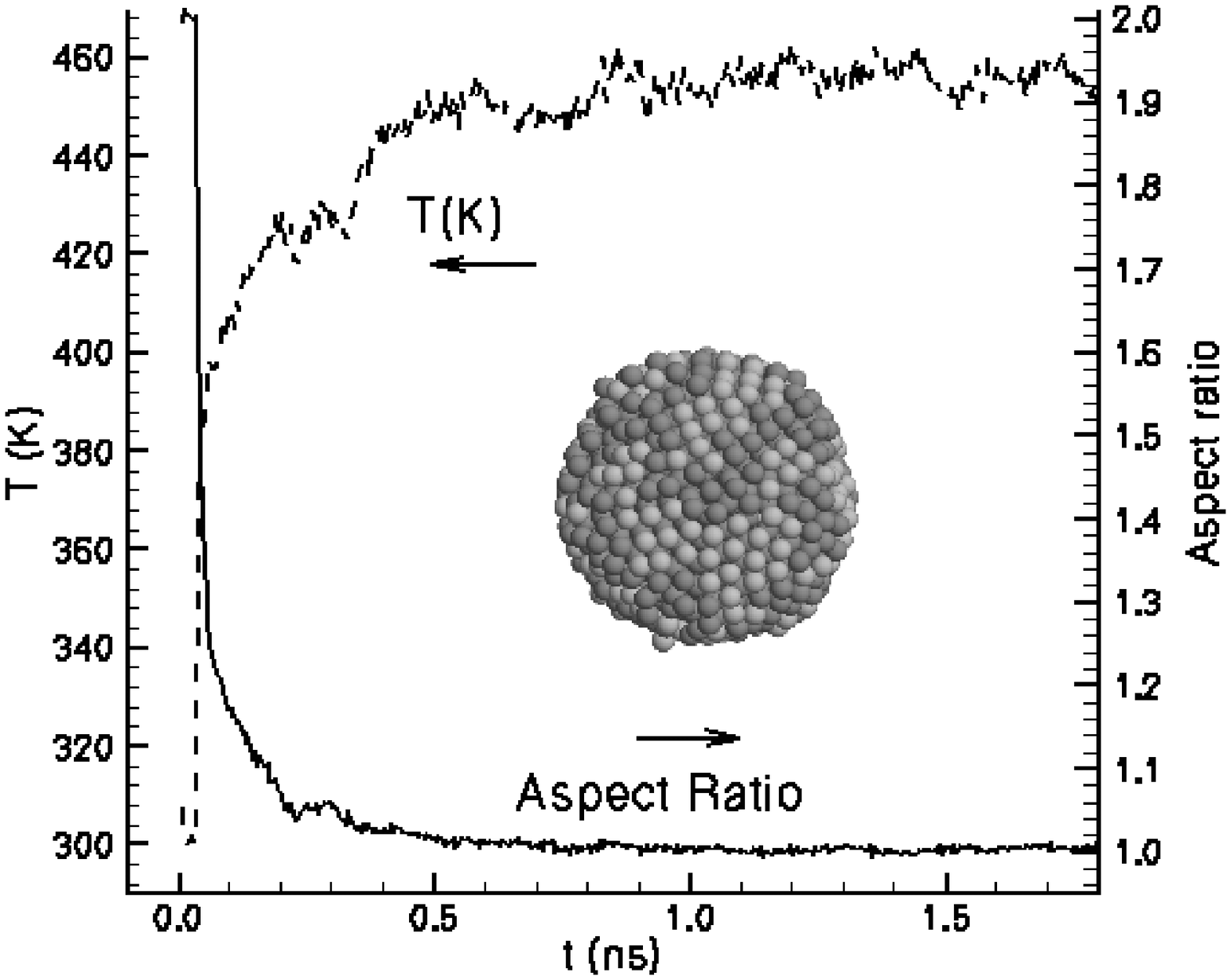}}
\caption{\label{temp-merge-ico} Evolution of the temperature and the aspect ratio during the coalescence of two
565-atom icosahedra at 300 K. At the point of first contact between clusters (approximately 30 ps into the
simulation), the temperature rises rapidly as surface energy is released. The inset shows the final cluster
structure.}
\end{figure}

The constant-energy simulation then proceeds at a total energy equal to the sum of the final energies from each isolated
equilibration run. Here we use a 3.75 fs time step which ensures energy is conserved to within 0.01 eV/atom over a 10 ns
simulation. Figure~\ref{frames} shows a series of snapshots of the initial stages of coalescence between two
565-atom clusters each equilibrated at 300 K. Coalescence typically begins tens of picoseconds into the simulation
when an atom on one of the aligned edges makes contact with the adjacent edge of the other cluster. The first snapshot
in figure~\ref{frames} shows the situation at this initial moment of contact, with the formation of an initial neck
shown in subsequent frames. Neck growth is essentially complete by the final snapshot in figure~\ref{frames}.
Figure~\ref{temp-merge-ico} shows the evolution of temperature and the aspect ratio
\footnote{Aligning the clusters along the x-axis, we define the aspect ratio to be $(R_y+R_z)/2R_x$ where $R_\alpha$
is the radius of gyration along axis $\alpha=x,y,z$.}

As the neck grows, the surface area of the merging clusters decreases, leading to the release of considerable
surface energy. This energy is converted to heat, as can be seen in Figure~\ref{temp-merge-ico} by the large
increase in temperature during neck growth. After initial contact and the period of rapid neck growth,
the aspect ratio evolves more slowly. In Figure~\ref{temp-merge-ico}, we can see that the cluster is essentially
spherical in shape after a period of 0.5 ns after initial contact, as the aspect ratio is very close to one by this
stage. Examination of cluster structure reveals that it is static after this time. We conclude that coalescence is
complete within 1 ns. The final structure of the 1130-atom cluster is that of a surface-reconstructed icosahedron
(see Figure~\ref{temp-merge-ico}).

\begin{figure}
\resizebox{\columnwidth}{!}{\includegraphics{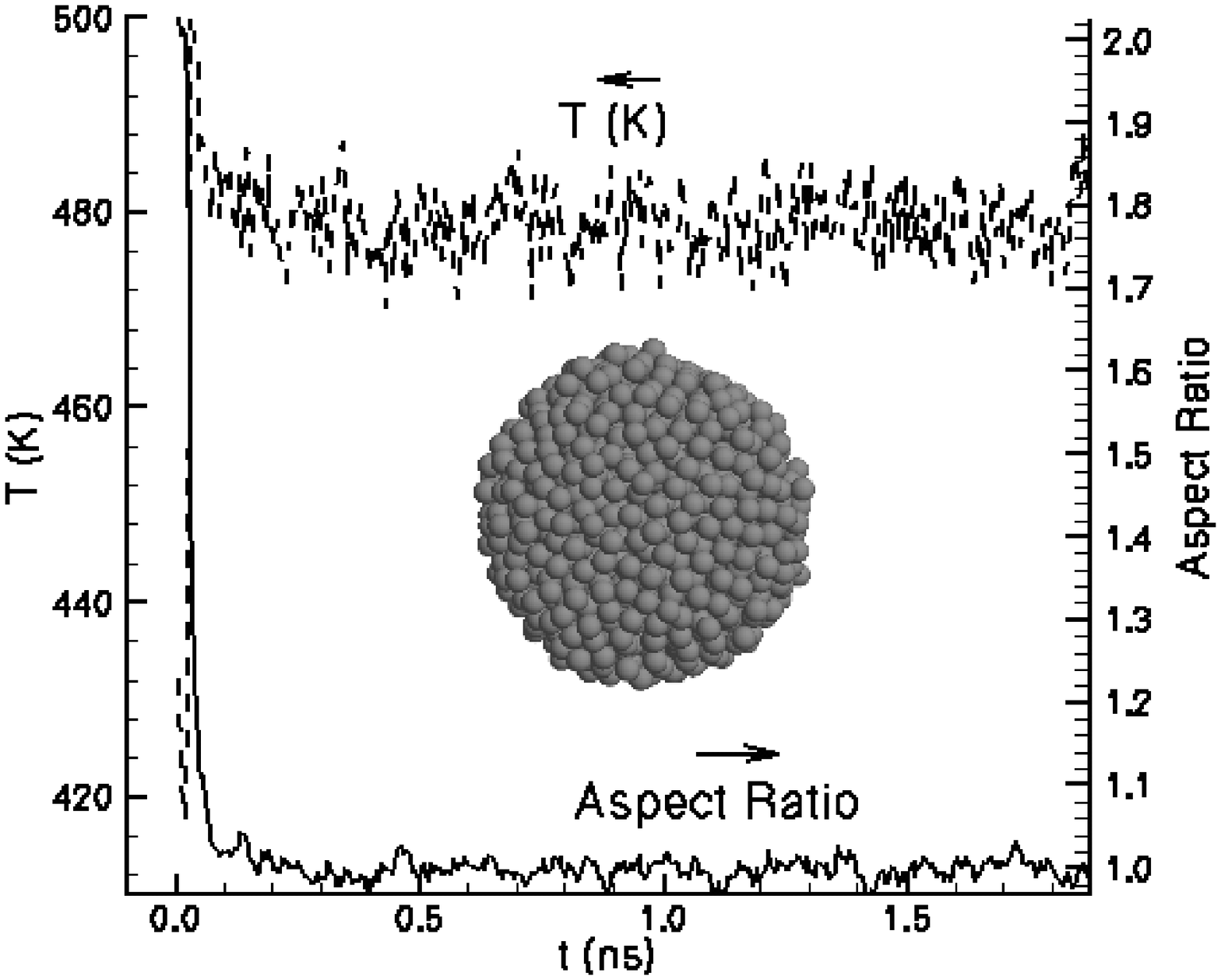}}
\caption{\label{mergemelt} Evolution of the temperature and aspect ratio during the coalescence of two
565-atom icosahedra initially at 420 K. At the point of first contact between clusters (approximately 30
ps into the simulation), the temperature rises rapidly as surface energy is released. The final state
of the coalesced cluster is a molten droplet shown in the inset.}
\end{figure}

The amount of surface energy released as heat will clearly play a role in enhancing atomic diffusion during coalescence.
In fact, if the initial temperature of the two clusters is close enough to the melting point, the final coalesced
cluster may be molten. Such a case is shown in Figure~\ref{mergemelt}, which illustrates the evolution of
temperature and aspect ratio of two clusters initially at 420 K. Here the coalescence is far more rapid than in
the previous example, occurring in tens of picoseconds rather than over 0.5 ns. The clusters form a neck and then
melt, with the coalesced cluster rapidly becoming spherical. Note the initial spike in temperature prior
to melting, followed by a drop in temperature due to the heat of melting. The final cluster is molten at a temperature of
approximately 480 K. In the IGA source, this coalesced cluster will cool slowly and freeze
before reaching equilibrium with the inert gas atmosphere.

It is instructive to consider simple models of coalescence and melting. Consider the coalescence of
a spherical particle of radius $R_1$ with a spherical particle of radius $R_2 \ge R_1$.
Conservation of energy tells us
that the increase in thermal energy of the coalescing cluster must balance the reduction in surface energy. Thus,
the increase in temperature of the aggregate can be estimated as:
\begin{equation}
\label{deltat}
\Delta T = {3 \sigma \over \rho c_v} {1 \over R_2} {(1+(R_1 / R_2)^2)-(1+(R_1 / R_2)^3)^{2/3} \over (1+(R_1 / R_2)^3)}%
\end{equation}
where $\sigma$ is the surface tension, $c_v$ is the heat capacity and $\rho$ is the density of the
two particles. In general, $\sigma$, $c_v$ and $\rho$ will depend on cluster size. However, we will ignore this
dependence here and use the bulk values for these parameters. For the simulation presented in
figure~\ref{temp-merge-ico}, $R_1 = R_2 = R(565) = 1.6$ nm and equation (\ref{deltat}) predicts $\Delta T = 160$ K, which
is close to the value attained in the simulation of 150 K. This increase in temperature is maximized when
the ratio of change in surface area to volume ratio is maximized i.e. when $R_1 = 0.852 R_2$ for a given $R_2$. 
If $T_i + \Delta T$ exceeds the cluster melting point then we would expect the cluster to melt
during coalescence.

However, it is well-known that the melting point, $T_M(R)$, of a metal cluster depends strongly on the radius, $R$.
A simple model for this dependence can be written as \cite{Pawlow09}:
\begin{equation}
\label{melt}
T_M (R) = T_B \left( 1-{L \over R} \right)
\end{equation}
where $L$ is some length which depends on bulk material parameters. Obviously, this model fails for clusters of
radius $R \sim L$ and does not account for the irregular variation in melting point observed in small clusters
\cite{Kusche99},
but experimental results for lead \cite{BenDavid95}, where $L \sim 0.5$ nm, suggest it is adequate for
$R \geq 2$ nm. By simulation we have identified the melting point of the
initial 565-atom clusters to be $T_M(565) = 400 \pm 10$ K and that of the coalesced 1130-atom clusters to be
$T_M(1130) = 480 \pm 10$ K. Equation (\ref{melt}) predicts the melting point for the 565-atom clusters ($R=1.6$ nm)
to be 410 K and for the 1130-atom clusters ($R=2.0$ nm) to be 450 K. This rough agreement between (\ref{melt})
and simulation will be sufficient for our purposes here.

Clearly one can always arrange an initial temperature $T_i$, such that $T_i+\Delta T > T_M((R_1^3+R_2^3)^{1/3})$.
However, for clusters of equal size $R_1 = R_2$, comparison of equation (\ref{deltat}) and equation (\ref{melt}) 
suggests that there is a critical cluster size $R = 1$ nm, below which coalescence will always cause melting 
(i.e. since at $T_i = 0$, $\Delta T > T_M((R_1^3+R_2^3)^{1/3})$). However, we have not tested this using simulations 
here as the glue potential \cite{LOE92} may be unsuitable for such small clusters \cite{Doye03}.

\begin{figure}
\resizebox{\columnwidth}{!}{\includegraphics{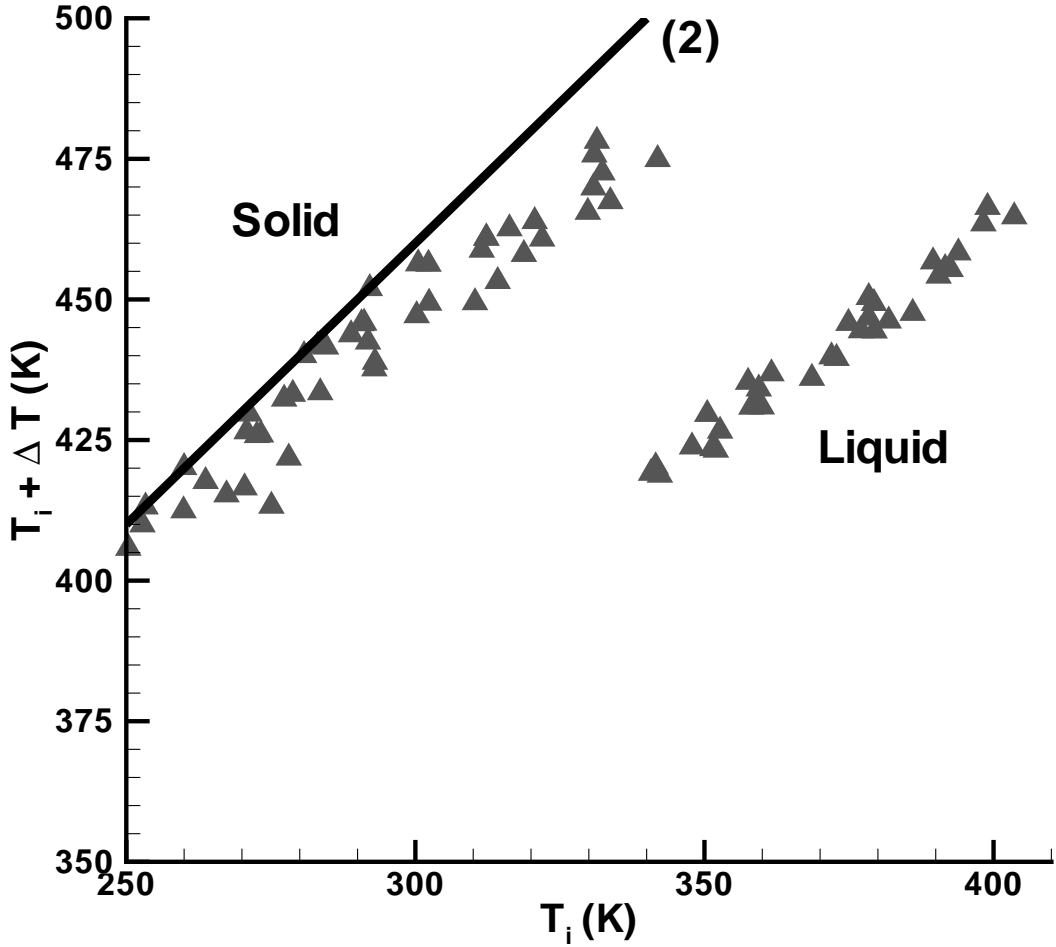}}
\caption{\label{ensemble} The final temperature of 75 pairs of coalescing 565-atom clusters as a function of initial
temperature. The solid line shows the final temperatures predicted by equation (\ref{deltat}). At initial
temperatures of 340 K and above the final coalesced cluster is molten and the final temperatures are lower
due to the latent heat of melting.}
\end{figure}

Figure~\ref{ensemble} shows the final temperatures of a series of pairs of 565-atom clusters that have undergone
coalescence. The initial temperature of the pairs has been varied from 250 K to 400 K. The solid line is the value
of $\Delta T$ predicted by equation (\ref{deltat}). We see from the figure, that equation (\ref{deltat}) is adequate
(considering the approximations involved) until an initial temperature of $T_i = 340 K$. Above this initial temperature
the final state of the coalesced clusters is molten. The highest final temperature reached in this series of
simulations is $T_i+\Delta T = 478 K$. Thus when the melting point of the final cluster is exceeded during coalescence,
a molten cluster results. The drop in the final temperature is due to the latent heat of melting. Note, that as the
initial temperatures of the simulations correspond to the total energy, figure~\ref{ensemble} is essentially a caloric
curve for the coalesced clusters. The coexistence of both liquid and solid clusters between final temperatures of
425 K and 475 K corresponds to the van der Waals loop in the microcanonical ensemble \cite{Wales94}.

\begin{figure}
\resizebox{\columnwidth}{!}{\includegraphics{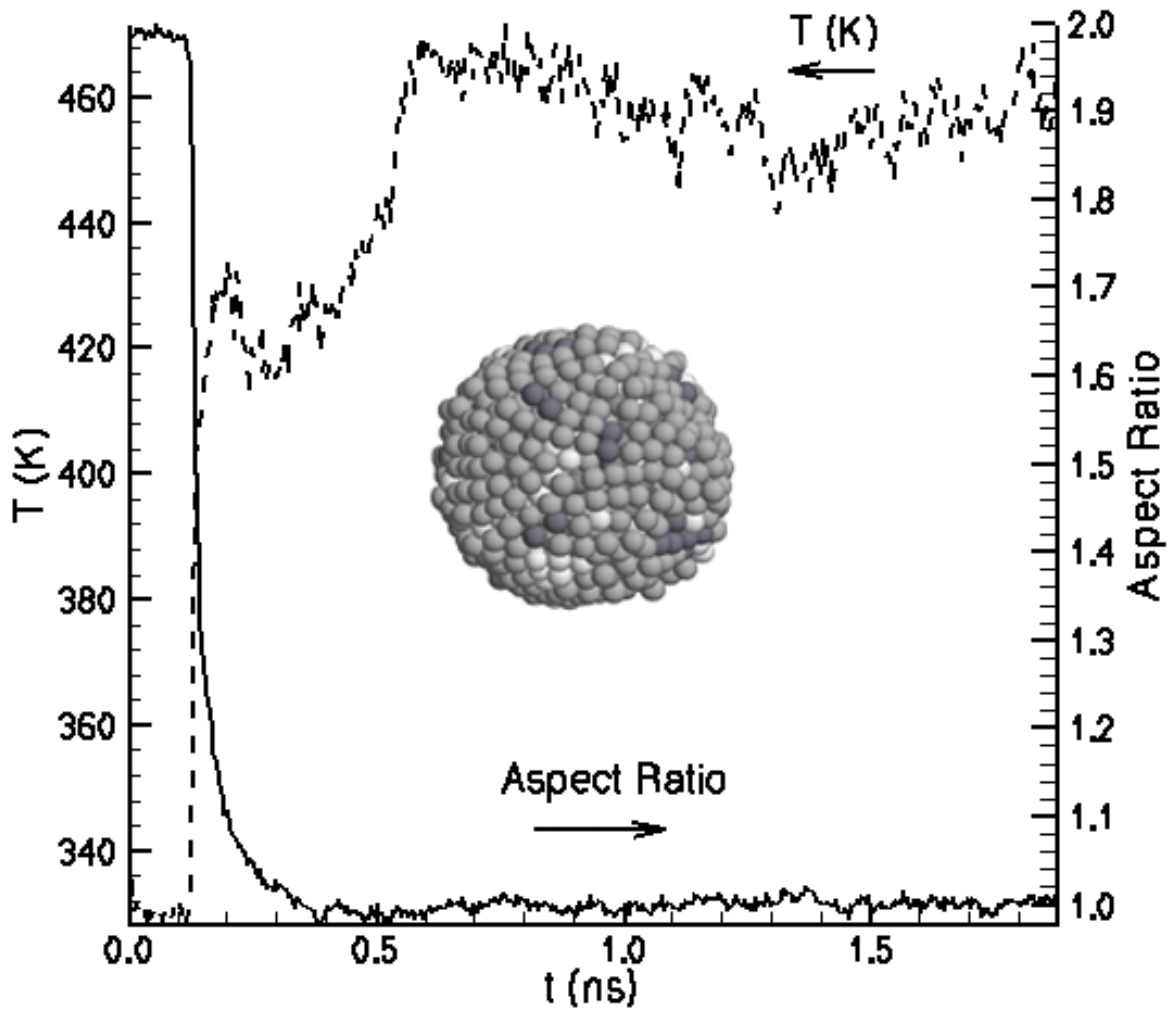}}
\caption{\label{melt-freeze} Evolution of temperature and aspect ratio during the coalescence of two 565-atom icosahedra
initially at 330 K. The initial coalesced structure is liquid and evidently unstable. The cluster subsequently
undergoes a structural transition to form an icosahedral structure. The inset shows the final cluster structure.}
\end{figure}

In figure~\ref{melt-freeze} we illustrate the coalescence of two 565-atom icosahedra initially prepared at 330 K.
Coalescence occurs rapidly, with the aspect ratio reaching one 0.2 ns after initial contact. There is a corresponding
increase in temperature of approximately $\Delta T \sim 90 K$ at this time. Inspection of the cluster structure at
this time reveals that it is highly disordered and is consistent with that of a molten structure \footnote{In fact,
it is difficult to be certain whether this transient state is liquid or a disordered glassy structure.}. The smaller
than expected increase in temperature during coalescence also suggests the cluster is molten. However this structure
is evidently unstable as at $t \sim $ 0.5 ns, the cluster structure undergoes a transition to form a more regular
icosahedral structure. There is a corresponding increase in temperature at this time. This route to coalescence via an
unstable molten intermediate state was found to be typical of coalescing clusters at initial temperatures
of 320-330 K. It is interesting to note that the temperature, $T=420$ K, of the unstable droplet falls between
$T_M(565)$ and $T_M(1130)$. This suggests a crude interpretation of this process, namely that the two individual
565-atom clusters melt when $T$ exceeds $T_M(565)$ during coalescence, but that the resulting 1130 atom molten droplet
is now unstable as $T < T_M(1130)$. Finally, it seems reasonable to characterize the liquid state seen immediately
after coalescence as metastable: while for small clusters dynamical coexistence between the solid and liquid state
has been observed \cite{Honeycutt87}, the energy barriers between the liquid and solid states scale
as $N^{2/3}$ \cite{Lynden-Bell}, hence repeated melting and freezing of the 1130-atom cluster will not be observed.

To summarize, we have identified three regimes for the coalescence of free solid metal clusters:
1. solid-solid coalescence by diffusion. This will produce highly-defective clusters, perhaps
with large numbers of stacking faults (see Ref.~\cite{Lewis97}). 2. complete melting of solid clusters as they
coalesce to form a single liquid droplet. In an IGA source, the final structures will
depend on the cooling rate due to equilibration with the inert gas \cite{Baletto02}.
3. complete melting of solid clusters coalescing to form an unstable liquid droplet which subsequently
solidifies. The solid structures produced by this mechanism will resemble those produced by a rapid quench.

\end{document}